\documentclass[sigconf, nonacm=true]{acmart}
\usepackage{color, colortbl}
\usepackage{bbding}
\usepackage{enumitem}

\usepackage[utf8]{inputenc}

\usepackage{graphicx}
\usepackage{subfig}

\usepackage[title]{appendix}

\AtBeginDocument{%
  \providecommand\BibTeX{{%
    \normalfont B\kern-0.5em{\scshape i\kern-0.25em b}\kern-0.8em\TeX}}}

\begin{document}

\title{A New Doctrine for Hardware Security}

\author{Adam Hastings}
\email{hastings@cs.columbia.edu}
\affiliation{%
  \institution{Columbia University}
  \city{New York}
  \state{New York}
}

\author{Simha Sethumadhavan}
\email{simha@cs.columbia.edu}
\affiliation{%
  \institution{Columbia University}
  \city{New York}
  \state{New York}}


\begin{abstract}
    In this paper, we promote the idea that recent woes in hardware security are not because of a lack of technical solutions but rather because market forces and incentives prevent those with the ability to fix problems from doing so.
    At the root of the problem is the fact that hardware security comes at a cost;
    Present issues in hardware security can be seen as the result of the players in the game of hardware security finding ways of avoiding paying this cost. 
    We formulate this idea into a doctrine of security, namely the Doctrine of Shared Burdens.
    Three cases studies---Rowhammer, Spectre, and Meltdown---are interpreted though the lens of this doctrine.
    Our doctrine illuminates why these problems and exist and what can be done about them.

\end{abstract}




\maketitle

\section{Introduction} \label{sec:intro}

% Outline
% Lots of recent HW problems

Once niche and arcane, the field of hardware security has recently become one of the most pressing issues in cybersecurity. 
Physical-level attacks like Rowhammer gave attackers the ability to modify a system's memory at will \cite{kim2014flipping}.
And microarchitectural side channel attacks like Spectre and Meltdown have shown how pervasive, dangerous, and hard-to-fix a hardware attack could be \cite{kocher2019spectre, lipp2018meltdown}.
Especially concerning is that these problems, while well-known and publicized, have generally not been fixed. Why?

% We have technical solutions, so what's up?

% We find that security is a burden, and that problems are not being resolved because people don't want to pay the cost
% Because of market failures (see appendix)
% We find that security is a burden, and that problems are not being resolved because people don't want to pay the cost
% Because of market failures (see appendix)
The answer, perhaps surprisingly, is not a lack of technical solutions.
Instead, we find that hardware security problems persist because suffers from a series of market failures such as information asymmetry, prisoners dilemmas, and markets for lemons, which disincentivize those who are able to fix serious security vulnerabilities from doing so (see Appendix A for more on market failures in hardware security).
Underpinning these market failures is the fact that hardware security usually comes at a cost in terms of performance, power, or area. 
We propose the notion that the poor state of hardware security is due to the various agents in the game of hardware security trying to avoid paying this cost.

We crystallize this notion into a conceptual framework called the Doctrine of Shared Burdens, which we present in Section 2.
Our doctrine also illustrates why prior doctrines of security do not apply to the domain of hardware security in Section 3. 
We then use the Doctrine of Shared Burdens to illuminate some of the most serious problems in hardware security in recent years, namely Rowhammer, Spectre, and Meltdown.
We find that our Doctrine of Shared Burdens incisively reveals the true issues behind these troublesome vulnerabilities and explains why have persisted or why they arose in the first place.
Section 3 also uses our doctrine to shed light on how researchers, engineers, and policymakers can work to fix them.
Finally, this paper concludes in Section 4.

% Hardware security, like other types of security, is an abnormal good, where the economic laws of supply and demand do not necessarily produce the most efficient outcome.
% As we will demonstrate, hardware security succumbs to many of the classic market failures such as information asymmetry,  free riding, misaligned incentives, and markets for lemons.
% We can understand the persistence of serious hardware vulnerabilities as a series of market failures, where market forces do not properly incentivize those who are able to fix the problems from doing so.

% We propose the Doctrine of Shared Burdens, which frames HW security as a cost paid by different people
% The doctrine is a conceptual framework for thinking about future and existing problems and who should go about fixing them
% We analyze three case studies.

% \input{motivation.tex}

\section{ Doctrine of Shared Burdens} \label{sec:doctrine}


\begin{figure*}[t]
    \begin{center}
    \begin{minipage}{.45\textwidth}
      \centering
      \includegraphics[width=.9\linewidth]{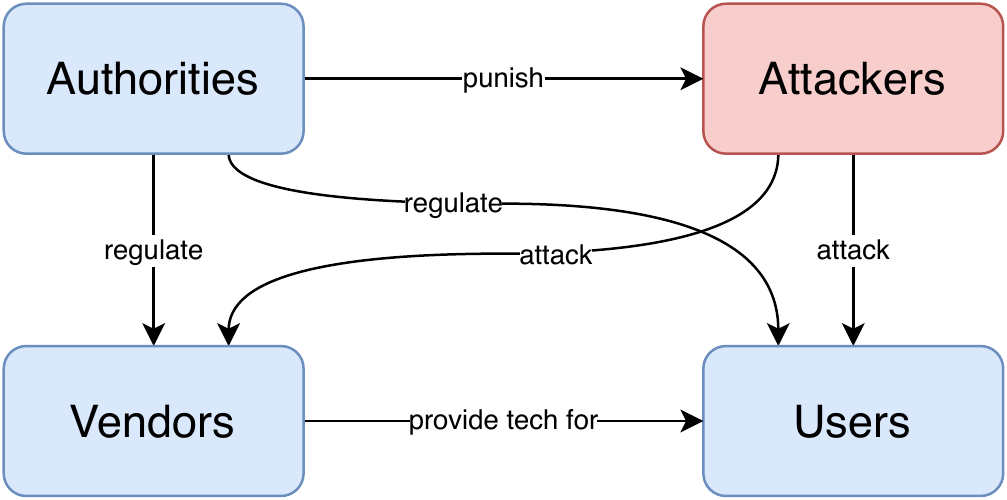}
      \captionof{figure}{The four players in the game of hardware security, and the relationships between them. The Doctrine of Shared Burdens says that systems should be designed so that the burden of security is distributed between these four players.}
      \label{fig:test1}
    \end{minipage}%
    \qquad
    \begin{minipage}{.45\textwidth}
      \centering
      \includegraphics[width=.9\linewidth]{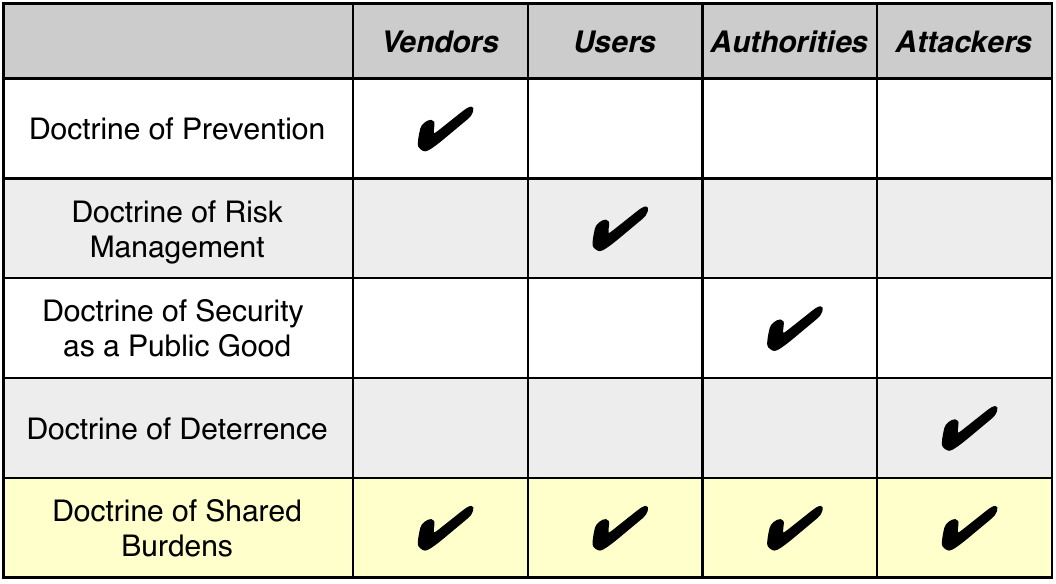}
      \captionof{table}{Prior doctrines focus on assigning the burden of security to a single party. Our new proposed doctrine of shared burdens argues that security must be seen as a burden to shared among all the players in the game of security.}
      %         \label{tab:my-table}}
      \label{tab:tab1}
    \end{minipage}
\end{center}
\end{figure*}

We propose a doctrine of hardware security based on the premise that hardware security is a burden that comes with a cost. 
This cost is necessarily borne by at least one of the four players in the game of hardware security, namely the Vendors, Users, Authorities, and Attackers.

\subsubsection{The Vendors}
The Vendors are the agents who design and and build systems for profit.
% The Vendors encompass the entire stack of computing systems, from the CPU vendors to the DRAM and flash memory vendors to the device manufacturers to the operating system writers and the system applications developers.
% The Vendors include the CPU and DRAM designers as well as the semicondudctor foundries and device manufacturers.
In our doctrine, the Vendors must bear the burden of ensuring that their products are safe and not easily exploitable. 
The Vendors pay their burden to security through the cost of validating their products against vulnerabilities, as well as through the opportunity cost of not making products that are more competitive in the marketplace but less secure.

\subsubsection{The Users}
% Users != victims
% Defenders are those who own the equipment that is hacked
% Targets?
% "Victim" makes sense, but our "victims" are not always those who actually suffer
%
% The defenders are those who are trying to defend assets from Attackers.
%\subsubsection{Maintainer}
%The maintainers are those who manage and administer technology for others. 
%In most organizations this includes the system administrators, who must consistenly install patches, configure firewalls, and manage access to their respective networks.
%The maintainers also include the Internet Service Providers, who Users are those who control the equipment that is hacked by the Attackers. 
The Users are the victims of attack.
% \footnote{Some may point out that those who are hacked (i.e. the Users in our model) are not always the same as those who actually suffer most from insecurity, and that the latter should be considered the ``true'' victim. For example, if a company is breached and its customers' credit card numbers are exposed, the customers may suffer financial losses whereas the company itself may only suffer minor loss of reputation. However, this is immaterial to the argument being made here: Regardless of who actually suffers from insecurity, defense is the responsibility of those who actually control and operate the systems.}.
The scope of who can be considered a ``User'' can range significantly, from a smartphone user to the cybersecurity team at a large organization. 
In our doctrine, it is the responsibility of the Users to secure their systems as best as possible.
Importantly, this means that a User is responsible for the protection of the assets they are entrusted with.
% Defenders can pay for their burden of security in many ways
% But how...?
% TODO I need help filling out this section
%The defenders must share some of the burden of security.
% The defenders should bear the responsibility of defending the assets placed under their protection.
% One way the makers pay their burden of security is through the cost of validating their products against vulnerabilities.

The Users pay for security by incurring the always-on cost of defending their systems.
The Users must also uphold their responsibility to security by applying patches and by not disabling security features, or else the Users will end up free riding off of the security efforts of others.

\subsubsection{The Authorities}
% Coercive ability
The Authorities are the regulatory bodies that have a degree of authority over the Vendors and the Users. 
The Authorities have the unique ability to correct the failures in the marketplace for security.
The role of the Authorities is often assumed by governments, but not always.
For example, self-regulatory organizations (SROs) are non-governmental regulatory groups which have a degree of regulatory authority over certain industries
\footnote{A well-known and exemplar SRO is the Financial Industry Regulatory Authority, Inc. (FINRA) which regulates and arbitrates all stock market operations in the United States. FINRA is a non-governmental organization comprised of the very members it regulates. FINRA is authorized by the U.S. Securities and Exchange Commission to enforce the rules and regulations of the the securities industry.}
;
SROs and other non-governmental agents can assume the role of an Authority as well.
An Authority has the burden of regulating, mandating, and sometimes enforcing the Vendors and the Users to uphold their respective responsibilities to security.
This can come in the form of mandates or regulations, e.g. a mandate that Vendors use two-factor authentication in their products.
An Authority also has the responsibility of punishing and prosecuting the Attackers when possible.

\subsubsection{The Attackers} 
Finally, the fourth player in the security game of our shared-burden doctrine is the Attacker. 
The Attacker is the party who perpetrates cybercrime.
% At first, this may seem odd that the Attacker is one of the security game players that must share the burden of security---After all, we can't expect a criminal to play by the rules set by the other three players.
% We include the Attacker not as a player who must uphold their responsibility of security to the others, but as a player who  pays the price for their decision to attack security.
% It is a different kind of burden or cost from the other three players, but a cost nonetheless.
Our doctrine posits that we should make attacks as expensive as possible for the Attackers in an attempt to discourage them from attacking systems in the first place.

We view the ``cost'' an Attacker pays as the opportunity cost they incur by choosing to attack a system, as well as the consequences they may face as a result of this decision.
We can consider the opportunity cost to be the amount of effort or resources an Attacker may need to expend to be successful. 
Our doctrine dictates that we should build defenses that require a high cost for the Attacker to overcome. 
Importantly, our doctrine also says that we should do so without overburdening the User, who must pay for the always-on cost of defense.
In other words, our defenses should be deliberately asymmetric against the Attackers and should offload the ``cost'' of security from the Users to the Attackers as much as possible.
One such example is cryptography, which imposes a minor overhead cost to Users with the proper keys, but an enormous cost for an Attacker to break if they don't possess the necessary keys.
Finally, we view the risk an Attacker assumes when breaking the law as another form of ``cost'' the Attacker pays.
Our doctrine says that we should make Attackers pay for their actions by improving the Authorities' ability to catch and prosecute the Attackers and  increasing the risks and punishments of attacking.
Both of these costs paid by the Attackers are intended to discourage and deter attacks.

\section{Differences from Prior Doctrines} \label{sec:rw}

We find that existing doctrines of  cybersecurity doctrines are insufficient because they misplace the burden of security among the players in the security game or do not apply to the domain of hardware security.
% Ultimately produce lower quality security and at a higher cost.
We examine four doctrines of security promulgated by Mulligan and Schneider \cite{mulligan2011doctrine}.

\subsection{Doctrine of Prevention} 
The Doctrine of Prevention states that security should be achieved by eliminating bugs and vulnerabilities from systems.
While a worthwhile goal,
Mulligan and Schneider point out that such a doctrine is impractical to achieve.
First, even a vulnerability-free system can still be overcome by social engineering or insider malfeasance.
Second, even if we ignore the human element, the computational cost of proving a design to be free of vulnerabilities is often impractical \cite{mulligan2011doctrine}.
Mulligan and Schneider position this in terms of software security but the arguments largely hold for hardware security as well, which is why most hardware designs are not formally verified.
However, even a formally verified design does not guarantee a secure product, as formal proofs rest on assumptions about how a design will be used and the environment under which it will be operated, and are only valid if these assumptions are met.
% For example, Sawada and Hunt formally verified a relatively advanced processor \cite{sawada2002verification},
% but an attacker can still hack such a system by getting it to violate one of its assumptions or by targeting something that the proof considers out-of-scope, such as side channels.

We add that the Doctrine of Prevention also fails because it assumes that the Vendors are sufficiently motivated to build secure products in the first place.
Failures in the marketplace for hardware security show that this is not always the case. 
The Doctrine of Shared Burdens clarifies that an Authority is needed to regulate and sometimes coerce the Vendors into creating secure products.

\subsection{Doctrine of Risk Management} 
The Doctrine of Risk Management takes a more pragmatic approach to  security by acknowledging that vulnerabilities and attacks are inevitable.
Rather than trying to build perfectly secure systems, the Doctrine of Risk Management puts security into terms of probabilities and expectations.
According to this doctrine, security administrators should prioritize finding and fixing the vulnerabilities that are 1) most likely to be exploited, and 2) most likely to cause harm if exploited.
This doctrine posits that security ought to be seen as an investment against future attacks and financial losses, and that the ``right'' level of security is whatever is best for an organization's bottom line.

% Mulligan and Schneider present this doctrine from the perspective of software or network security.
This doctrine fails because there is a lack of accurate and publicly available information on threats and attacks, making it very difficult to quantitatively reason about the risks of cyber attack and build useful actuarial models.
The consequence is that security practitioners are rarely able to make metric-driven decisions on how to best secure their systems, and instead must resort to an ad hoc and qualitative approach to this doctrine.

% The Doctrine of Risk Management doesn't work for hardware security, and not just because the risks are hard to measure.
% Hardware security is effectively limited by what the Vendors are willing to do to fix the vulnerabilities in their products.
% Consider the microarchitectural data sampling (MDS) bugs found in Intel processors in 2019 \cite{ridl,canella2019fallout, schwarz2019zombieload}.
% The recommended fix is to disable Hyper-Threading, which is a performance technique that lies at the root of the MDS vulnerabilities.
% However, Intel has decided that MDS is not a serious enough risk to warrant disabling Hyper-Threading. 
% We note that the performance cost of disabling Hyper-Threading could be as much as 40\% for certain workloads~\cite{applemds}.
% In other words, Intel has assumed the role of deciding the what level of risk is tolerable on behalf of its customers, denying Users the ability to decide this for themselves.
% Furthermore, even if Users are concerned about security, in general they don't have the ability to configure what they deem to be the appropriate level of risk for themselves or their organizations.

The Doctrine of Risk Management also falls short in its allocation of the burden of security.
The doctrine requires the Users to assume the full cost of security.
Market failures and inefficiencies promise that this strategy will always be suboptimal, as Users are subject to free riding and typically don't know enough about security to prevent markets for lemons. 
The doctrine underburdens the Vendors, who are free to sell products known to be insecure, and it underburdens the Attackers, who face no repercussions for their actions.

\subsection{Doctrine of Deterrence} 
The third doctrine of cybersecurity that Mulligan and Schneider highlight aims to discourage crime by improving authorities' ability to catch and prosecute the Attackers.
This doctrine is hard enough to achieve in software and network security---cybercrime forensics are limited in their ability to assign blame, mostly because our current internet infrastructure is poorly equipped to handle attribution.
This can be partially offset with robust logging, but prosecution remains difficult as it often crosses international borders.

The problem becomes even more challenging in the domain of  hardware security.
Recent microarchitectural attacks such as Spectre \cite{kocher2019spectre} and Meltdown \cite{lipp2018meltdown} demonstrated exploits that are essentially silent to the User. 
How can a User catch a cybercriminal red-handed if the User has no way of telling that they are being attacked in the first place? 
The lack of threat of prosecution means that there is little to deter the Attackers.

Even if better attribution were possible, this doctrine fails because it aims to put the entire cost of security onto the Attacker, and ignores the responsibilities the Authorities, Vendors, and Users must play in achieving security.
Rather than leaving the security holes open and prosecuting the Attackers later, it would be more efficient for an Authority to hold the Vendors more accountable and require them to close the security holes in the first place.

\subsection{Doctrine of Cybersecurity as a Public Good} 
% Mulligan and Schneider propose the doctrine that cybersecurity is non-rivalrous and non-excludable, making it a public good by definition. 
Mulligan and Schneider propose viewing security as a public good using another well-studied public good---public health---as an exemplar.
According to Mulligan and Schneider, it is the responsibility of an Authority (namely the government) to ensure and administer public health, through activities such as public education, disease prevention, and disease control.
Public health is a mature model of how and where an Authority's obligation to protect the population can supersede individual liberties.
Using this as a framework, Mulligan and Schneider define the goals of the public goods cybersecurity doctrine as (i) providing public cybersecurity, and (ii) managing insecurity in a way that balances individual rights and public welfare.

This doctrine improperly assigns the burden of security to the Authorities alone.
% Along this line of thought, the Doctrine of Cybersecurity as a Public Good was called into question by Weber \cite{weber2017coercion}.
% Weber argues that to apply the public health model to cybersecurity may require a level of Authority coercion far beyond what the society is currently willing to accept.
As a result, applying the public health model to cybersecurity may require a level of Authority coercion far beyond what the society is currently willing to accept \cite{weber2017coercion}.
For example, we see little precedent for something like a  government-enforced cyber-quarantine or cyber-vaccinations, and have little reason to believe a government can (or should) take on the full responsibility of security.

We add that the sheer complexity of hardware is also a severe hindrance to effective Authority-administered security.
Only the Vendors and industry experts know how best to secure their products;
Authority intervention would inevitably be heavy-handed and misguided.
Allowing the Vendors to secure their own devices, but holding them liable for their products' security would be more efficient.
The role of the Authorities should be to regulate industries and correct market failures, but not to administer security wholesale.

This doctrine also falls short in its framing of the problem, as it fails to hold the Attackers responsible. 
The doctrine lacks the notions of punishment and deterrence. 
% For example, communicable diseases such as viruses cannot be punished for their ``actions'', and are not deterred by the threat of prosecution.
A full-picture view of security needs to consider the Attacker as an active participant in the struggle for security, and particularly as a self-interested participant who is motivated by the rewards of hacking but deterred by its drawbacks, such as the real or perceived risk of being caught and punished. 
% A complete doctrine of security needs to include the notion of deterrence in one way or another.

%\subsubsection{Laissez-Faire}
%The laissez-faire doctrine says that 

% \section{Applying the Doctrine to Hardware Security}
\section{Case Studies} \label{sec:cases}

% How does our doctrine help with existing hardware security problems?

We examine three recent high-profile problems in hardware security---Rowhammer, Spectre, and Meltdown---as case studies that illustrate how our doctrine can inform us on how we should allocate the burden of security.
% We now use our doctrine of shared burdens to illuminate some of the open problems facing hardware security today.

\subsection{Rowhammer}

% The DRAM industry's response to the rowhammer problem, from its initial discovery to what seems to be a viable solution (RFM), shows how the burden of security can be allocated among the four players of the security game.
% This case study also provides examples of how the thorny non-technical challenges can be overcome, namely how responsibility over a community-wide problem can be allocated and distributed in a self-regulatory body such as JEDEC.
% Finally, it is worth pointing out that JEDEC, a non-governmental organization with no coercive power, was able to get its members to agree to collectively take on the burden of securing their products against rowhammer.
% The takeaway is that we shouldn't overlook the role that community standards can play in achieving security.
% Perhaps other tough, community-wide security problems can be solved if we leverage standardization to make it in the best interest of those who have the ability to fix longstanding problems to do so.

% In 2014, the DRAM industry was rocked by the revelation
% of the rowhammer vulnerability. Kim et al. discovered that
% repeated, persistent access of a DRAM wordline could cause
% enough of an electrical disturbance to change the bits stored
% in neighboring DRAM cells \cite{kim2014flipping}.
% The DRAM industry's response to the rowhammer problem shows how the burden of security can be allocated among the four players of the security game.
% The response also shows how the thorny, non-technical challenges of a vulnerability can be overcome via collaborration and standardization.

Rowhammer is a problem found in modern Dynamic Random Access Memory (DRAM), the technology behind main memory in virtually all computing devices.
We use it as a case study because it exemplifies an end-to-end application of our Doctrine of Shared Burdens, from initial discovery to what we believe to be its solution. 
It provides an excellent model of how the burden of defense can be distributed between Vendors, Users, Authorities, and Attackers.

\subsubsection{The Attack}

DRAM is a victim of its own success.
For the last forty years, its transistors have shrank tremendously, allowing for an exponential increase in density (bits stored per unit area).
Rowhammer is an unintended consequence of this tremendous density.

As DRAM cells (essentially just a transistor and a capacitor, capable of storing a single bit) got smaller and smaller, two things happened.
First, they became more delicate and more susceptible to losing the data they stored.
And second, as DRAM components became more tightly packed together, they started electromagnetically interfering with each other.
In 2014, it was shown that this interference could be reliably harnessed to alter the contents of the data stored in DRAM by repeatedly accessing the same row of DRAM memory (henceforth known as ``hammering''), wherein the fluctuations in voltage on the DRAM's internal wires could flip the values of the bits stored in nearby DRAM cells \cite{kim2014flipping}.
And in 2015, this primitive was demonstrated to enable a working exploit \cite{seaborn2015exploiting}.

% In fact, a modern DRAM device can have several billion transistors on a single chip, all very small and tightly packed together.

% During this time, we've seen an exponential decrease in the size of transistors and an exponential increase in the density of DRAM chips, to the point where there are now many billions of transistors on a single DRAM chip.
% Rowhammer is an unintended consequence of this tremendous density.

Rowhammer is a serious hardware vulnerability because it breaks basic integrity guarantees in computer systems by allowing Attackers to modify unauthorized memory locations,
enabling an entire new class of attacks.
Rowhammer has demonstrated dangerous potential, and can be leveraged to achieve privilege escalation \cite{seaborn2015exploiting}, cross-VM attacks \cite{xiao2016one}, and even as a side channel to read privileged data \cite{kwong2020rambleed}. 
To make matters worse, there's no easy fix---bits can
flip faster than a doubled DRAM refresh rate can fix, and
are more numerous than error correcting codes (ECC) can correct \cite{kim2014flipping}.

\subsubsection{Who Should Fix Rowhammer?}

How the DRAM industry has handled the rowhammer problem is an interesting and illuminating case study in how overall security is often a function of the distribution of burdens between the Vendors, the Users, the Authorities, and the Attackers. 
We begin by looking at the balance of burdens immediately after the rowhammer problem was identified.
Since ECC couldn't fix the problem, the only available rowhammer defense a User could employ was to increase the DRAM refresh rate\footnote{DRAM is \textit{dynamic}, meaning that the DRAM cells will lose their contents unless they are periodically refreshed.}, thus reducing how long a DRAM row could be ``hammered'' before being automatically refreshed. 
Unfortunately, simply doubling the refresh rate doesn't fix the problem---the authors estimate that, in the worst case, the refresh rate would need to increase by a factor of \textit{seven} in order to fully mitigate bit flips \cite{kim2014flipping}.
Since refresh is already such a expensive operation (in terms of both DRAM latency and energy), the overhead of such a defense would come at a tremendous cost.
By increasing the refresh rate, it is ultimately the Users who pay for security, who suffer from slower memory that consumes more energy.
Through the perspective of our doctrine of shared burdens, we see that the cost of security is placed solely on the Users, and that the other players are not shouldering their fair share of the burden.

Rowhammer is a flaw in DRAM products, and should be the responsibility of the Vendors to correct.
% Any true solution to the rowhammer problem requires the Vendors to take on some of the burden of security themselves. 
However, it is not immediately clear \textit{which Vendor} should be responsible for fixing the problem:
For DRAM to be used, it requires an external memory controller---typically located on the same chip as the CPU---to issue commands such as reads, writes, and refreshes. 
And of the various rowhammer defenses promoted after the  vulnerability became known, some advocated for rowhammer to be fixed by the memory controller whiles others advocated that the problem should be fixed within the DRAM chips themselves.
Since the DRAM chips and memory controllers are made by different companies, who should be responsible for fixing the problem?

If the memory controllers vendors and DRAM vendors operated completely unconstrained, it is reasonable to believe that neither side would voluntarily take on the burden of fixing rowhammer.
Each side could rightfully claim that it is the responsibility of the other side to fix the problem.
However, the memory controller vendors and the DRAM vendors do \textit{not} operate wholly unconstrained. 
Both sides belong to JEDEC, a DRAM industry trade organization.
JEDEC decides and defines standards for DRAM technologies, including the interface between DRAM devices and memory controllers, 
which the JEDEC members must then follow.
Importantly, JEDEC members don't join because they like being told how their products should behave; 
Rather, JEDEC members join because it is in their own self interest to do so:
Standardization increases cross-compatibility between DRAM and the devices that use it, effectively opening up a DRAM vendor's products to a wider consumer base.
For a DRAM vendor \textit{not} to comply with JEDEC standardization would essentially be a death sentence, as no memory controller vendor would 
want their product to be reliant on a single DRAM vendor. 
We see then that JEDEC has a high degree of authority over the DRAM industry, and can act in the role of an Authority in our doctrine of shared burdens.

\begin{center}
    \noindent{\fbox{\parbox{0.9\linewidth}{\textbf{Takeaway \#1:} Trade organizations and standards committees can fulfill the role of an Authority.}}}
\end{center}

But JEDEC is a non-governmental organization, and is not guaranteed to make decisions that are in the best interest of widespread security.
After all, JEDEC is comprised of self-interested companies;
If these companies collectively decided against a standardized rowhammer defense, it is plausible that rowhammer would remain unsolved.
Indeed, if JEDEC was comprised solely of DRAM and memory controller vendors, this might be the case. 
But in addition to DRAM and memory controller vendors, JEDEC also contains a significant number of DRAM \textit{consumers}, in industries ranging from cloud computing and automotive to aerospace and defense.
Since these consumers, who play the role of the User rather than the Vendor, stand to lose more in case of DRAM insecurity than the Vendors, it is likely that their presence in JEDEC has influenced the standardization committees towards seriously addressing the rowhammer problem.

\begin{center}
    \noindent{\fbox{\parbox{0.9\linewidth}{\textbf{Takeaway \#2:} Consumer (User) interest groups may be needed to motivate an Authority into acting on their behalf.}}}
\end{center}

\subsubsection{Attempt \# 1: TRR}

One of the first concerted efforts to address rowhammer
problem was  Targeted Row Refresh (TRR), an optional mode of operation defined in JEDEC's 
2014 LPDDR4 standard (the fourth generation of low-power
DRAM) \cite{lpddr4}. And while not part of the DDR4 (fourth generation DRAM, intended for higher-performance applications
than LPDDR4) standard, some DRAM vendors opted to include
TRR in some of their later DDR4 products as well. 
TRR wasn't a prescriptive order telling DRAM vendors how to fix
rowhammer, but was more of a contract between memory controller and DRAM device on the high-level actions that should be taken in the presence of excessive row activations. 
Patents give hints on how each vendor may have internally implemented TRR  \cite{patent1, patent2, patent3, patent4}, but ultimately the architecture used commercial devices is largely unknown.
While TRR was a first step towards rowhammer protection,
it unfortunately was doomed to fail---it could only refresh
a limited number of DRAM rows per DRAM refresh window, and was only designed to refresh the rows physically
adjacent to excessively activated (i.e. hammered) rows. This
allowed rowhammer attacks with multiple targets to overwhelm TRR's ability to fix the hotspots, and did not address
the issue of rowhammers affecting more than just the nearest
physically adjacent row \cite{frigo_trrespass_2020}.
Given that the LPDDR4 standard
was released only a few months after the rowhammer bug was
announced, it is not surprising and perhaps even expected that
the DRAM industry's first attempt at fixing the rowhammer
problem was not without flaw.

Despite its shortcomings, one of the DDR4 vendors (Vendor C in the TRRespass paper) seems to have leveraged TRR
successfully enough to completely protect against many variants of rowhammer \cite{frigo_trrespass_2020}. 
However, because vendors A and B also implement TRR yet still fall susceptible to rowhammer, we must conclude that the difference between success and failure is not the protocol defined in the standard itself, but the private, proprietary, and vendor-dependent architecture used to implement TRR.
In short, the LPDDR4 standard alone does not offer sufficient protection against rowhammer. 
JEDEC was presumably aware of this, which is why they introduced a new rowhammer defense in the next generation of DRAM.

\subsubsection{Attempt \# 2: RFM}

In early 2020, JEDEC released the LPDDR5 standard.
In LPDDR5, JEDEC replaced
TRR with Refresh Management (RFM) \cite{lpddr5}. 
In RFM, the DRAM device counts activate commands per bank, and issues a refresh once a threshold number of activations has been reached.
RFM requires a degree of coordination between the memory controller and DRAM device, and distributes the burden of defense between the DRAM vendors and the memory controller vendors.
While LPDDR5 devices are just now entering the market and researchers have not yet been able to experimentally evaluate RFM, the tightness of the standard suggests that RFM will provide a very high level of protection against rowhammer attacks, possibly even eliminating the rowhammer problem altogether.  
And while the DDR5 standard hasn't yet been released, we suspect that it will include a very similar if not identical rowhammer defense.

RFM is an asymmetrical defense that punishes the Attacker but not the User.
In the LPDDR5 standard, RFM is defined to have the memory controller count the number of times a row of DRAM memory is accessed via an \texttt{ACT} (activate) command.
If the number of activations exceeds some threshold, the memory controller issues a special type of refresh command (\texttt{RFM}) 
which applies fine-grained, selective refreshes to ``hot'' regions in the DRAM (essentially regions of DRAM cells that have repeatedly been accessed since the last time the region was refreshed).
The DRAM chips are specially designed to account for this special type of refresh and still maintain performance.
Therefore, if an Attacker tries to rowhammer a region of DRAM by repeatedly activating rows in some region of DRAM, RFM will automatically refresh the victim region, nullifying any advances the Attacker had made.
The standard is flexible enough to allow the DRAM and memory controller vendors to ``tune'' the defense so that the RFM mechanisms spring into action before an Attacker is able to make any headway in flipping DRAM bits.
Yet at the same time, the RFM standard is designed not to burden the User (the DRAM owner and user) with excessive, additional refreshes, minimizing the always-on cost of rowhammer defense.
Then if RFM is properly implemented, we expect that it will successfully shift the burden of security away from the User and onto the Attacker.
RFM only burdens systems with its full weight in the presence of anomalous behavior.

\begin{center}
    \noindent{\fbox{\parbox{0.9\linewidth}{\textbf{Takeaway \#3:} The burden of security can be offloaded from the User to the Attacker by punishing anomalous behavior.}}}
\end{center}

The DRAM industry's response to the rowhammer problem, from its initial discovery to what seems to be a viable solution (RFM), shows how the burden of security can be allocated among the four players of the security game.
This case study also provides examples of how the thorny non-technical challenges can be overcome, namely how responsibility over a community-wide problem can be allocated and distributed in a self-regulatory body such as JEDEC.
Finally, it is worth pointing out that JEDEC, a non-governmental organization with no coercive power, was able to get its members to agree to collectively take on the burden of securing their products against rowhammer.
We see that standards can play a huge role in achieving security by getting Vendors to agree to take on some of the cost of security.
Perhaps other community-wide security problems can be solved if we leverage standardization to make it in the best interest of those who have the ability to fix longstanding problems to do so.

%%%%%%%%%%%%%%%%%%%%%%%%%%%%%%%%%%%%%%%%%%%%%%%%%%%%%%%%%%%%%
%%%%%%%%%%%%%%%%%%%%%%%%%%%%%%%%%%%%%%%%%%%%%%%%%%%%%%%%%%%%%
%%%%%%%%%%%%%%%%%%%%%%%%%%%%%%%%%%%%%%%%%%%%%%%%%%%%%%%%%%%%%

\subsection{Spectre} 

Another serious open threat to hardware security is Spectre \cite{kocher2019spectre}.
Announced in early 2018, Spectre demonstrated that speculative execution, a performance-enhancing feature in modern processors, could be exploited in a dangerous new type of attack.
Unlike Rowhammer, Spectre is still very much an open problem that has little in the way of usable and deployable solutions.

\subsubsection{The Attack}

Spectre targets speculative execution.
Speculative execution in processors can be defined as any action that is taken preemptively and on the expectation (but not guarantee) that a program's execution will follow one path and not some other path.
Modern processors employ many types of speculation as an effective means of improving performance.
In the canonical Spectre attack, the type of speculation targeted was branch prediction, which works as follows:
When a processor is executing a program, it frequently encounters branches in the execution path.
Branches can come in many forms.
They can be \textit{conditional}, where a particular execution path is taken if some set of conditions are met and not taken if the conditions are not met (an \texttt{if} statement is a simple example of this).
Branches can also be \textit{indirect}. 
Unlike conditional branches, which explicitly tell the CPU the address at which to start executing if the condition holds, an indirect branch instead tells the CPU where the address is located.
As it turns out, both types of branches are highly predictable using on-line machine learning algorithms built into the hardware. 
CPUs can achieve dramatic performance improvements if they can correctly predict branch direction, because they can start speculatively executing the branch before the program's actual branch direction is known.
If the branch prediction was correct, then the speculatively executed instructions are confirmed to be correct, and the program is further along in its execution than if the CPU has waited until the direction of the branch was known.
But if the branch prediction was wrong, the speculatively executed instructions are incorrect, and must be purged from the CPU.

It is in this way that Spectre takes advantage of speculative execution. 
Spectre maliciously mistrains the branch predictor to purposely mispeculate and access out-of-bounds data.
Before Spectre, such mispeculation wasn't thought to be a security problem, 
% because the once the CPU realizes the branch was mispredicted it invalidates the speculated instructions.
because the CPU invalidates speculated instructions once it realizes it mispredicted.
But Spectre showed that the results of mispeculated, out-of-bounds instructions could be exfiltrated even when invalidated.  
Spectre exfiltrates data through cache timing side channels such as Flush+Reload \cite{gullasch2011cache, yarom2014flush} or Evict+Reload \cite{gruss2015cache}, although other microarchitectural side channels could theoretically be used as well. 
While originally demonstrated to attack branch prediction, Spectre attacks can target many of the types of speculation found in modern CPUs.

Spectre is a serious threat to security at large.
It has the potential to read arbitrary memory locations, including cryptographic keys.
But perhaps the most troubling to security researchers is that there is no real solution to the problem.
Computer architects have not yet found a way to engineer themselves out of the problem, and there is no widely accepted solution.

\subsubsection{Who Should Fix Spectre?}

The only fail-safe defense currently available is to turn off speculation altogether. 
This is comparable to the early days of the rowhammer vulnerability, where the only available solution was to raise the refresh rate to intolerable levels.
Much like the rowhammer problem was several years ago, the burden of defense against Spectre rests far too heavily on the Users of systems, and not nearly enough on the Vendors, Authorities, or Attackers.
Spectre defenses will require a rebalancing of these roles and responsibilities before a solution can be achieved.

As with rowhammer, we can first look to the Vendors to take upon more of the burden of security. 
Unlike rowhammer, which had multiple parties partially to blame, with Spectre we know exactly who needs to fix the problem: the CPU vendors.
Unfortunately, there is no JEDEC-like organization among CPU vendors to standardize what a defense should look like. 
% No standards in (this domain)
In the terms of our doctrine of shared burdens, there is no Authority that can exert its authority and get the CPU vendors to build Spectre defenses into their products.
Likewise, there are no consumer advocate groups that have enough influence to motivate the CPU vendors to fix the problem either.
Finally, information asymmetry prevents the market from correcting the problem:
Due to the sheer complexity of CPUs, consumers won't be able to evaluate and properly price the value of a Spectre defense, and won't pay a premium for a feature (Spectre security) they can't identify, causing a market for lemons.
This puts the CPU vendors in a prisoner's dilemma.
It would be beneficial for all if the vendors cooperate and agree to jointly fix Spectre in their products, but the threat of defection from competing Vendors makes this an irrational proposition.
In other words, it is rational for the CPU Vendors to \textit{not} fix Spectre, at the expense of security as a whole.
Unconstrained and financially unmotivated, 
% and without anyone or group fulfilling the role of the Authority, 
we shouldn't expect the CPU vendors to take upon the burden of fixing Spectre without first applying some kind of exogenous pressure.
% No standard to include timing in ISA, so side channels seem like they will always be around

\begin{center}
    \noindent{\fbox{\parbox{0.9\linewidth}{\textbf{Takeaway \#4:} Spectre stands to remain unresolved, because the CPU vendors are stuck in a prisoner's dilemma and there is no Authority to correct this market failure.}}}
\end{center}

Even if a JEDEC-like organization did exist for CPU vendors and was able to coordinate a standardized defense, what would such a defense look like?
% First, the members of such an organization would have to agree to implement the defense and would have to commit to not defect from this agreement; 
% Otherwise the CPU vendors are placed into a prisoner's dilemma where defection is the rational choice.
% Second, t
The defense would have to be specific enough to fix the problem but general enough to allow for variations in implementations between the different Vendors.
One way we can foresee such an approach would be a tax on performance or energy.
This approach balances the burden between Vendors and Users, with the Vendors paying to implement the defense and the Users paying (in terms of performance or energy) for the always-on overhead.

In our search for a solution to Spectre, we must also consider the balance of the burden of security between the Attacker and User.
For defenses to be accepted by the Users, we need the always-on, recurring cost of defense to be tolerably low.
Likewise, in accordance with our doctrine of distributed burdens, we want the burden of defense to be asymmetrically placed onto the Attackers.
We would like defenses that can flare up when anomalous behavior is detected.
We can consider mispeculation to be the anomalous behavior.
However, since mispeculation is a regular occurrence even in benign program execution, we need a defense that doesn't needlessly punish programs for mispeculating.
We can look to the adaptive lockout mechanisms on phones and laptops as an example:
Perhaps the ``punishment'' meted out by a Spectre defense should scale with the number of mispeculations within some time frame, where punishment could be something like the speed at which a CPU allows a process to execute.

%%%%%%%%%%%%%%%%%%%%%%%%%%%%%%%%%%%%%%%%%%%%%%%%%%%%%%%%%%%%%
%%%%%%%%%%%%%%%%%%%%%%%%%%%%%%%%%%%%%%%%%%%%%%%%%%%%%%%%%%%%%
%%%%%%%%%%%%%%%%%%%%%%%%%%%%%%%%%%%%%%%%%%%%%%%%%%%%%%%%%%%%%

\subsection{Meltdown}

Meltdown is a hardware-based attack that was announced at the same time as Spectre \cite{lipp2018meltdown}. 
Meltdown shares some similarities with Spectre, and uses some of the same mechanisms as the Spectre attack, and therefore is sometimes conflated with Spectre or thought to be a rooted in the same underlying problem.
However, while superficially similar, Spectre and Meltdown are fundamentally different problems and must be thought of as such. 
Seeing Meltdown through the lens of our doctrine requires its own interpretation, independent of Spectre.
From this case study we see an example of where market forces can incentivize a Vendor to fix some types of hardware security problems \textit{without} the need for an Authority, and examine the circumstances that make this possible.
% The unique qualities of Meltdown make it an interesting case study for our doctrine.

\subsubsection{The Attack}

Meltdown is a consequence of an optimization found in many  processors.
The optimization (and its deleterious effects)  stem from the way that some processors handle faults.
A fault is an exception raised by the hardware when code tries to do something unallowed or undefined, such as a division by zero, or in the case of Meltdown, an attempt to read privileged kernel data from an unprivileged process. 
When such a fault occurs, the processor will typically halt or kill the offending process.
Prior to Meltdown, the way many processors handled faults would best be described as ``lazy'', meaning that the they wouldn't deal with the faults for as long as possible.
More specifically, vulnerable processors wouldn't kill such an unauthorized memory access until just before the faulting instruction retires and updates the architectural state of the program.
Presumably, some CPU vendors chose to build their fault handling this way because it enabled some kind of optimization somewhere else in the processor.
At first glance, this seems like a perfectly reasonable thing to do---eventually the fault is caught, and before the faulty access is able to update the program, so where's the harm?
The problem is that between the illegal memory access and the exception being raised, for a brief period of time the unauthorized memory access resides somewhere in the processor's microarchitectural state. 
Meltdown is a way of exfiltrating this secret memory value in the small time window between unauthorized access and fault handling. 

Meltdown leverages a performance-enhancing technique known as \textit{out-of-order execution} to exfiltrate the secret value.
Out-of-order execution is a performance-enhancing technique used in processors wherein instructions are allowed to execute as soon as their operands are available rather than being required to wait to execute in program order, and is permitted insofar as the instruction reordering still preserves program correctness.
In a Meltdown attack, malicious out-of-order instructions use the secret value obtained by the unauthorized memory access \textit{before} the exception is handled, which can affect microarchitectural structures such as the L1 data cache. 
Like Spectre, Meltdown then uses a cache side channel timing such as Prime+Probe or Flush+Reload to leak the secret.

In the canonical Meltdown attack, the target is kernel memory. 
Kernel memory is typically mapped into the virtual address space of every process as a performance enhancement technique---it allows for kernel memory pages to remain in memory and in the translation lookaside buffer (TLB) when the operating system undergoes a context switch.
Since Meltdown provides a way for an unprivileged process to read privileged data, all of kernel memory becomes readable.
And because the kernel itself typically contains the virtual-to-physical mappings of all of physical memory, it becomes possible to read any memory location from inside any unprivileged user space process.

\subsubsection{Who Should Fix Meltdown?}

Clearly, Meltdown is a serious problem in desperate need of a solution.
What is not immediately clear is \textit{how} it should be fixed, and who should pay for the cost of defense.
Like Rowhammer, there are multiple parties who could implement a solution to Meltdown. 
We now use our doctrine of shared burdens to help us understand the problem and how to fix it.

% Our doctrine doesn't illuminate anything. We're just cherry picking examples...
% The makers are responsible for security, but the Users and Attackers should pay the cost.

Let's first look at the defense originally proposed by the Meltdown authors: KPTI \cite{gruss2017kaslr}.
Kernel page table isolation (KPTI, also formerly known as KAISER) actually predates Meltdown, as it was originally intended to solve another problem, namely a kernel side-channel attack against KASLR (kernel address space layout randomization), itself a defense against memory safety exploits in the operating system's kernel.
As it turns out, KPTI defends against Meltdown as well.
KPTI essentially removes the kernel from each processes' address space, thus denying Meltdown its attack surface.
KPTI had previously been deployed as a Linux patch, and was implemented in Windows and OS X  patches during a responsible disclosure period before Meltdown was announced. 
Despite the patches Meltdown was not completely fixed, as KPTI still leaves a residual attack surface.
It also came with a hefty performance overhead for Users, who were stuck paying (in terms of performance) to defend a product that was initially advertised as secure.
Like Rowhammer and Spectre, the first available solution was costly to the Users and allows the Vendors to avoid responsibility.

% Initial patch unfair to OS vendors, as it wans't their problem
% Was it fair to burden the operating system vendors
Another problem with relying on KPTI to fix Meltdown is that it places the burden of defense on the operating system vendors, who were suddenly asked to fix a problem they didn't create.
This is unfair, as the Doctrine of Shared Burdens says that Vendors should be held responsible for the security of their own products.
Upon a close examination of Meltdown, it is very clear that the problem does not come from the OS vendors but rather the CPU vendors:
Meltdown is possible because some CPU vendors bypassed a security domain (privileged data accesses from unprivileged processes), which is a violation of an architectural security principle. 
In other words, Meltdown was not an out-of-the-blue, completely unexpected and unprecedented attack like Spectre;
Rather Meltdown may best be described as simply a bug, and clearly the CPU vendors should be held responsible.

% Clearly the CPU vendors should be held repsonsible

% Takeaway
\begin{center}
    \noindent{\fbox{\parbox{0.9\linewidth}{\textbf{Takeaway \#5:} Fixing Meltdown is the responsibility of the CPU vendors whose products were insecurely designed.}}}
\end{center}

% Prisoner's Dilemma? No, market forces in x86 fix it
Since the CPU industry lacks an Authority that can set rules and mediate problems, we may expect a prisoner's dilemma  similar to Spectre that prevents CPU vendors from fixing the problem.
However, Meltdown arose under certain circumstances that has allowed the free market to partially fix the problem on its own.
We highlight two circumstances---endogenous to the marketplace---that have helped fix Meltdown. 
First, at least in the x86 marketplace which dominates desktop and server computing, Meltdown was isolated to only one CPU vendor---Intel---while its main competitor, AMD, was unaffected.
And second, Meltdown (and Spectre) received an enormous amount of publicity at the time, unprecedented for a hardware vulnerability.
Meltdown was covered by major mainstream news organizations, and it became known far outside the niche domain of hardware security.
This undoubtedly broke down the information asymmetry between Intel and its consumers, who now knew of a problem that while they maybe didn't fully understand, were definitely aware that Intel's products were vulnerable.
Consumers then could then knowingly choose between a Meltdown-susceptible processor or a Meltdown-free processor.
Clearly, it was in Intel's best financial interest to fix their processors as fast as possible to make them more competitive in the marketplace.
And in late 2018, that's exactly what Intel did.
Intel announced its new Whiskey Lake architecture, which among other things, had in-silicon fixes to the problem.

% Takeaway
\begin{center}
    \noindent{\fbox{\parbox{0.9\linewidth}{\textbf{Takeaway \#6:} Market forces can sometimes fix problems on their own without the need for a coercive Authority.}}}
\end{center}

% Caveat: Information Asymmetry, time to announce defense
While this may seem like a success of the free market, there are some notable caveats that need to be addressed. 
The primary issue is that without an Authority to mediate vulnerability disclosure, there are tremendous incentives for Vendors to delay known vulnerabilities and downplay their risks once they are known. 
We see this in the case of Meltdown and particularly in the later but related MDS attacks \cite{ridl, canella2019fallout}.
In both cases, the vulnerability was known to CPU vendors (in this case, Intel) for a very long time---over a year in the case of the MDS attacks---before the vulnerability was publicly disclosed.
This is very different from software security, where the process from vulnerability discovery to patch is typically 90 days or less.
This tremendous delay between vulnerability discovery and vulnerability defense hugely exacerbates the information asymmetry between Vendor and User, as the Vendor is selling a known insecure product to the User without the User's knowledge, potentially for many months if not longer.
% How to fix?
% Researchers should go to SRO to mediate
There was an even larger gap between when Intel first learned of Meltdown and when they first announced their intent to fix it.
To fix these problems, the intervention of an Authority may be needed.

We propose the use of a self-regulatory organization (SRO) to act as an Authority and improve vulnerability response.
Such an SRO needs to be comprised of the members it regulates, who are the only ones who understand the sheer complexity of modern hardware designs and how to best regulate them.
Under this approach, vulnerability researchers would no longer have to wait on the Vendor's terms before announcing discovered vulnerabilities, and would no longer have to try to talk the Vendors into fixing the discovered problems.
A SRO could act as a mediator between vulnerability researchers and Vendors, and could wield the authority necessary to bring the Vendors to make meaningful change.
% Takeaway

\begin{center}
    \noindent{\fbox{\parbox{0.9\linewidth}{\textbf{Takeaway \#7:} Authorities such as SROs need to mediate the disclosure of vulnerabilities.}}}
\end{center}

\section{Conclusion} \label{sec:conclusion}

The prognosis for hardware security is grim.
The rate at which serious vulnerabilities are discovered far outpaces the computer hardware industry's ability (or motivation) to fix the problems.
% We expect that the severity of the problem will only increase without some sort of widespread action.
Without a common understanding of why these problems arise and persist we may never achieve true hardware security.
The Doctrine of Shared Burdens aims to rectify this by providing a conceptual framework that advances the discourse on what kinds of action need to be taken and where they should come from.

Implementing the Doctrine of Shared Burdens creates a unique set of challenges for engineers, researchers, and policymakers. 
For example, assuming an Authority decides to greater regulate Vendors' products, what kind of technical solutions enable an Authority to audit a Vendor's security mechanisms?
What kind of technical solutions enable Vendors to better punish anomalous behavior coming from the Attackers?
Likewise, what technical solutions enable Authorities to better catch and prosecute the Attackers?
How does an Authority ensure that Users are following sound security practices without infringing on User privacy?
How can Users trust that the products they buy have been designed to be secure?
Such questions (and their solutions) may be critical in fixing longstanding security problems and preventing new ones altogether.

% In the conclusion write  a sentence or two about how it would be useful to have technical solutions to implement fire safety model for hardware security including punishments, automatic “fire safety”  inspections,  etc.

% In this paper, we presented a new doctrine for hardware security that aims to advance the discourse on what these kinds of widespread actions should look like and where they should come from. 

% Without a conceptual framework for the decision makers and regulators to use, we run the risk of crafting policy that misappropriates the burden of security and misses the bigger picture of how security should be administered.

% \begin{acks}
% \end{acks}
  
\bibliographystyle{IEEEtran}
\bibliography{main}

\begin{appendices}

\section{Market Failures in Hardware Security}

A market failure is an economic situation wherein a free market fails to produce the most efficient distribution of goods or services.
Market failures are a known issue in security that have been discussed in the past \cite{anderson2001information}.
Despite hardware security's unique characteristics, we find that it succumbs to many of the same market failures as other areas of security.
In this appendix we examine four types of market failures from the perspective of hardware security.

\subsubsection{Information Asymmetry/Markets for Lemons} 
A common failure of open and free markets is information asymmetry. 
In this failure, one party of an economic transaction has more or better information than the other. 
For example, consider a scenario where a hardware company knows of serious security vulnerabilities in their product but decides that it would be too costly to fix.
It would be rational for a self-interested company to not publicly disclose the vulnerability for fear that it would damage their reputation and hurt sales.
There is then an imbalance or asymmetry of information (i.e. the presence of the vulnerability) between the company and its customers.
Without knowing of the serious vulnerabilities, customers will continue to buy the product to their own detriment.
Breaking down this information asymmetry would push customers to purchase safer, competing products instead, and would incentivize the company to patch the vulnerability.

Information asymmetry leads to a related market failure known as the market for lemons \cite{akerlof1978market}.
The market for lemons, first explained in the context of the marketplace for used cars, is a situation wherein information asymmetry degrades the quality of goods in the marketplace.
Imagine a marketplace of used cars, where some of the cars are of good quality  while others are defective (the ``lemons''). 
% Imagine a marketplace of hardware goods such as CPUs.
% Assume all CPUs in the marketplace offer the same performance, but some have built-in defenses against certain hardware attacks (e.g. Spectre) while others do not. 
The car dealers will price the cars accordingly, selling the more valuable good cars at a higher price point and selling the lemons for cheap.
% The CPU vendors, knowing which CPUs are secure and which CPUs are insecure, will sell the more secure (and more valuable) CPUs at a higher price and sell the insecure CPUs (i.e. the ``lemons'') at a lower price.
But only the car dealers know the difference between the good cars and the lemons, because the buyers don't know enough about cars to distinguish the good from the bad (because of  information asymmetry).
% But the consumers, who are not knowledgeable enough to put a 
% But there is information asymmetry in the market, and the consumers are 
% But there is information asymmetry in the market, and only the CPU vendors know the difference between a secure and insecure product, because the 
Buyers, not wanting to purchase a lemon, will be willing to pay a fixed price somewhere between the price of a good car and a lemon.
% Consumers, not wanting to purchase an insecure CPU, will be willing to pay a fixed price somewhere above the cheapest CPU on the market (since it is likely a ``lemon'') 
The result is that the dealers will only sell when they possess lemons, because they will be selling a low-value car at a price higher than the car is worth and make a profit. 
Likewise, dealers will leave the marketplace when they possess good cars, because the dealers won't want to sell a good car for less than it's worth.
This introduces a negative feedback loop that pushes the good cars out of the marketplace and floods it with low-quality, defective lemons.

In other words, customers are unwilling to pay a premium for a feature or quality they can't identify.
We see the same principle apply to hardware security.
Hardware is dizzyingly complex, and the average customer is not technically knowledgeable enough to be able to distinguish secure from insecure products.
The effect is that the more secure but more expensive products will be driven out of the marketplace, leaving behind only the cheaper, less-secure products.
Widespread security suffers as a result, and the general welfare of society is worse off.

\subsubsection{Prisoner's Dilemma} 
The prisoner's dilemma, borrowed from game theory, is a scenario where it is rational for two self-interested entities to not cooperate, even if it is in their best interest to do so. 
The classic formulation is as follows:
Assume there are two prisoners who are being charged for the same crime. 
In order to prosecute either prisoner, the prosecutors need the second prisoner to testify against the first.
If the two prisoners cooperate by remaining silent and refuse to rat each other out, then there isn't enough evidence to convict either prisoner, and so the two prisoners are only convicted of lesser charges and given a light prison sentence.
But the prosecutors, being clever, offer immunity to either prisoner if they testify against the other, and without being testified against themselves. 
This means that a prisoner who betrays the other or ``defects'' walks away free while their partner in crime is given a severe prison sentence.
If both prisoners betray the other (i.e. ``defect'') by witnessing against the other, then both prisoners are convicted and are given a moderate prison sentence.

The optimal outcome for the prisoners is achieved when the prisoners cooperate and don't testify against each other.
However, the incentives are such that it is rational for both prisoners to defect, making both prisoners worse off than if they had cooperated.
We can see similar dynamics in security. 
Imagine a market for lemons where products are insecure and the customers can't distinguish between safe and unsafe products (but can distinguish between product performance). 
It is in the best interest of all for the product vendors to agree on some kind of collective action (i.e. cooperate) to fix their products' insecurity.
Assume that this collective action towards defense will degrade the products' performance slightly but greatly enhance security.
The incentives are then the same as the prisoners in the prisoner's dilemma:
It is rational for a product vendor to defect (i.e. not implement the defense and not take the performance hit) in the hopes that the competing vendors \textit{don't} defect, so that the defecting vendor's product has a performance edge over its competitor's products and becomes more competitive in the marketplace.
If all the vendors were rational, no one would cooperate to jointly improve security, leaving everyone in a suboptimal state of security.
In an unconstrained market, such prisoner's dilemmas can disincentivize those with the ability to improve security from doing so.

\subsubsection{Misaligned Incentives} Another marketplace failure occurs when those who responsible for providing security are not those who suffer in case of insecurity.
For example, consider a CPU vendor with a hardware flaw that allows attackers to read privileged memory, such as cryptographic keys or sensitive financial information.
If a user is attacked, only the user suffers the consequences, and not the CPU vendor whose insecure product allowed the attack to happen in the first place.
Combined with the market for lemons principle, this means that the CPU vendor has little incentive to create a more secure product.

% \subsubsection{Tragedy of the Commons} One of the best-known market failures is the so-called tragedy of the commons.
% The canonical example is a group of cattle herders who share access to some parcel of land (the ``commons''). 
% The cattle herders will rationally add as many cows to the commons as possible, to maximize their profits.
% The commons will soon become overgrazed and worthless to all.
% The outcome is that the cattle herders are worse off than if they had not strictly acted in their own self interest.

% Similar problems occur in security.
% W

\subsubsection{Free Riding} 
% Similar to the tragedy of the commons is the problem of free riding.
Related to the prisoner's dilemma is the problem of free riding.
This market failure occurs when those who benefit from some shared resource rationally do not pay for them.
% Put another way 
A common example is vaccinations:
If a community is widely vaccinated against some disease, it would be rational for an individual to forgo the hassle of vaccination, because the individual is protected by the herd immunity provided by the vaccinated population (i.e. the individual ``free rides'' off of the efforts of others).
Of course, if everyone applied this thinking, no one would get vaccinated, herd immunity would disappear, and disease would spread.
We see similar problems in security. 
% To illustrate this, consider a scenario where 
% This problem arises 
% Botnets

Because security is so dependent on the cyber health of an entire community, it is rational for an individual not to pay their fair contribution towards community security and free ride off of others.
For example, if a CPU vendor releases a patch that hurts performance, it is rational for the CPU user to not apply the patch in the hopes that the other community members \textit{do} apply the patch, allowing the user to free ride off of the security benefits of others without themselves taking the hit to performance.  
This leads to a market failure where no one patches and the community is worse off than if the individuals had patched their systems.

  \section{Fire Safety as an Extended Analogy} \label{sec:analogy}

Our Doctrine of Shared Burdens is highly analogous to fire safety.
In developing our doctrine, we found it useful to think in terms of metaphor and analogy and use fire safety as a starting point.
Fire safety, much like security, is a full-system property, where all aspects of the system must be secured for the system as a whole to be secure. 
But unlike hardware security, fire safety has been developed and refined for literally thousands of years.
% Fire safety has been developed and refined for literally thousands of years.
It provides a mature model of how society can collectively and efficiently respond to a persistent threat, where the roles of those responsible for fire safety are well-established, effective, and generally uncontroversial. 
In contrast, computer security has only seen a few decades of effort in an ever-changing landscape of threats and vulnerabilities, and we haven't had the time or stability to reach a mature theory on what the roles are in security and who should bear responsibility.
% We need a better understanding of who should bear responsibility for the different aspects of security before we can craft sensible regulations or mandates aimed at correcting marketplace failures.

We can use the fire safety model to illuminate what roles there are in hardware security and how they each contribute towards overall security.
% We can use the fire safety model to illuminate the failings in the way security is administered today.
The first and most obvious observation is that security today places far too much of a burden on the end user.
Users are being placed into situations where they are powerless to defend themselves, but are entirely responsible for their own safety and well-being.
This is akin to building a house made of matches and then blaming the residents when the house burns down. 
The analogy continues: 
Because building residents typically don't know the fire code, they cannot themselves distinguish between safe and unsafe buildings.
The effect is that the market cannot price houses accordingly (since buyers are unwilling to pay a premium for a feature they cannot observe) and it becomes rational for the architects to design cheaper but less-safe houses.
We see the same effect in hardware security:
Consumers---who are generally unaware of the hardware vulnerabilities in today's products---are purchasing unsafe hardware because they don't know any better, and subsequently assume all risk if the hardware is hacked.
And because consumers are unwilling to pay for protection against a threat they don't know exists, it becomes rational for hardware vendors to not include costly security measures in their products.
Only unsafe products remain in the marketplace, which stabilizes in a suboptimal state.

If security today places too much of a burden on the end user, then a corollary is that it does not place enough of a burden on the architect.
In the example of the house made of matches, it is obvious that such a house should have never been built in the first place. 
Present-day hardware security has no such notion.
Hardware vendors are free to market and sell products with serious vulnerabilities to the unknowing public, who are too unaware of the problem to nudge the markets into fixing the problem.

% Fire codes are adaptive
Another takeaway from the fire safety model is that present-day security is seen as a personal issue and not as a community responsibility. 
Indeed, there is no equivalent of a publicly-funded firefighter in today's security landscape. 
The lack of collective response to security threats means that metaphorical fires are spreading freely among hosts:
Botnets, largely composed of insecure and unpatched devices, are a routine threat because no one bears responsibility to take them down.
And worms that distribute dangerous malware are a scourge of connected devices because once again, there is no organized defense to stop the spread.

A final, key takeaway from the fire safety model is that the system depends on clearly-defined responsibilities.
Nothing is allowed to slip through the cracks.
And if a new fire ``vulnerability'' is found (perhaps a material is discovered to be excessively flammable, or a new type of building floorplan gets too congested during building evacuation), then the fire code is updated to attribute responsibility somewhere. 
In using the fire safety analogy, a new security doctrine must allow for adaptable responsibilities, especially in a field which changes so frequently.

% \subsection{Where the Analogy Fails}

% Before we use the fire safety model as a springboard for developing our own doctrine of hardware security, we should first be aware of its limitations.
% After all, fire safety is not computer security, and no model is perfect.
Of course, no model is perfect.
A key difference between hardware security and fire safety is that 
fires are accidents\footnote{Arson is an exception, but irrelevant here}, whereas cyberattacks are not.
Unlike fire, attackers deliberately find the weakest points in a system and exploit them.
However, we argue that this distinction has no meaningful consequences.
Both fire and cyberattacks take the path of least resistance.
And in both cases, the best defense is to patch the known weak spots.
We conclude that intent is irrelevant, and that this seemingly important difference bears no effect on the analogy's applicability.

%Besides any other potential differences between the two domains, we must also be careful to not let the metaphor do the driving

% Models are just models
%In the words of ``A metaphor is not a doctrine and it is certainly not a theory.''
%As the saying goes, all models are wrong, but some are useful.

\end{appendices}

\end{document}